\documentclass[twocolumn,10pt,aps, pra, nofootinbib,superscriptaddress,showkeys,showpacs,longbibliography]{revtex4-1}
\usepackage{amsmath,amsfonts,amssymb,amstext,amscd,amsthm,dsfont}
\usepackage{physics}
\usepackage{color}
\usepackage[colorlinks, citecolor=blue]{hyperref}
\usepackage{amsfonts}
\usepackage[font=small,format=plain,labelsep=period,labelfont=bf,textfont=normal,singlelinecheck=false,justification=raggedright]{caption}
\usepackage{braket,tikz}
\usetikzlibrary{arrows}
\usepackage{soul,xcolor}
\usepackage{gensymb}
\usepackage[normalem]{ulem}

\usepackage[T1]{fontenc}

\usepackage{soul}
\usepackage{graphicx}
\renewcommand{\max}{{\rm max}}
\renewcommand{\min}{{\rm min}}
\usepackage{braket}
\usepackage{placeins}

\begin{document}
	
\title{Bounds on the dynamics of periodic quantum walks and emergence of the gapless\\ 
and gapped Dirac equation}

\author{N. Pradeep Kumar}
\affiliation{The Institute of Mathematical Sciences, C. I. T, campus, Taramani, Chennai, 600113, India}
\author{Radhakrishna Balu}
\affiliation{U.S. Army Research Laboratory, Computational and Information Sciences Directorate, Adelphi, Maryland 20783, USA}
\affiliation{Computer Science and Electrical Engineering, University of Maryland \\ Baltimore County, 1000 Hilltop Circle, Baltimore, MD 21250, USA}
\author{Raymond Laflamme}
\affiliation{Institute for Quantum Computing and Department of Physics and Astronomy, University of Waterloo, Waterloo N2L 3G1, Ontario, Canada}
\affiliation{Perimeter Institute for Theoretical Physics, Waterloo, N2L 2Y5, ON, Canada}
\author{C. M. Chandrashekar}
\email{chandru@imsc.res.in}
\affiliation{The Institute of Mathematical Sciences, C. I. T, campus, Taramani, Chennai, 600113, India}
\affiliation{Homi Bhabha National Institute, Training School Complex, Anushakti Nagar, Mumbai 400094, India}


\begin{abstract} 
We study the dynamics of discrete-time quantum walk using quantum coin operations, $\hat{C}(\theta_1)$ and $\hat{C}(\theta_2)$ in time-dependent periodic sequence. For the two-period quantum walk with the parameters $\theta_1$ and $\theta_2$ in the coin operations we show that the standard deviation [$\sigma_{\theta_1, \theta_2} (t)$] is the same as the minimum of standard deviation obtained from one of the one-period quantum walks with coin operations $\theta_1$ or $\theta_2$, $\sigma_{\theta_1, \theta_2}(t) = \min \{\sigma_{\theta_1}(t), \sigma_{\theta_2}(t) \}$. Our numerical result is analytically corroborated using the dispersion relation obtained from the continuum limit of the dynamics.  Using the dispersion relation for one- and two-period quantum walks, we present the bounds on the dynamics of three- and higher period quantum walks. We also show that the bounds for the two-period quantum walk will hold good for the split-step quantum walk which is also defined using two coin operators using $\theta_1$ and $\theta_2$.   Unlike the previous known connection of discrete-time quantum walks with the massless Dirac equation where coin parameter $\theta=0$, here we show the recovery of the massless Dirac equation with non-zero $\theta$ parameters contributing to the intriguing interference in the dynamics in a totally non-relativistic situation. We also present the effect of periodic sequence on the entanglement between coin and position space. 
\end{abstract}
\maketitle


\section{Introduction}  
\label{intro}

The quantum walk is a generalization of the classical random walk equivalent in a quantum mechanical framework\,\cite{Ria1958, Fey86, Par88, ADZ93, Mey96}. By exploiting the quantum interference in the dynamics, quantum walks outperform the classical random walk by spreading quadratically faster in position space\,\cite{Kem03, Ven12}. At certain computational tasks, quantum walks provide exponential speedup\,\cite{CDF03,CG04} over classical computation and are used as a powerful tool in most of the efficient quantum algorithms\,\cite{Amb07,MSS07, BS06, FGG08}. Both the variants, continuous-time and discrete-time quantum walks have been shown to be universally quantum computation primitive, that is, they can be used to efficiently realize any quantum computation tasks\,\cite{Chi09, LCE10}. With the ability to engineer and control the dynamics of the discrete-time quantum walk by controlling various parameters in the evolution operators, quantum simulations of localization\,\cite{Joy12, Cha12, CB15}, topological bound states\,\cite{KRB10, OK11}, relativistic quantum dynamics where the speed of light is mimicked by the parameter of the evolution operator\,\cite{Mey96, Fre06, CBS10, Cha13, MBD13, MBD14, AFF15, Per16}, and  neutrino oscillations\,\cite{MMC17, MP16} have been shown. The quantum walk has also played an important role in modeling the energy transfer in the artificial photosynthetic material\,\cite{Eng07, MRL08}.  Faster transport\,\cite{HM09}, graph isomorphism\,\cite{DW08}, and quantum percolation\,\cite{CB14, KKN14} are few other application where the quantum walk has found application.

Experimentally, controlled evolution of quantum walks has also been demonstrated in various physical systems such as NMR\,\cite{RLB05}, trapped ions\,\cite{SMS09, ZKG10}, cold atoms\,\cite{KFC09} and photonic systems\,\cite{PLP08, PLM10, SCP10, BFL10} making it a most suitable dynamic process which can be engineered for quantum simulations.  

Among the two variants of quantum walks, the dynamics of the continuous-time variant are described directly on the position Hilbert space using an Hamiltonian. The dynamics of each step of the discrete-time variant are defined on a Hilbert space composed of both, the position and particle Hilbert space using a combination of unitary quantum coin operation acting only on the particle space followed by a position shift operation acting on both, particle and position space. By exploring different forms of quantum coin and position shift operators in homogeneous\,\cite{NV00, CSL08}, periodic\,\cite{MK09}, quasiperiodic\,\cite{XQT14, GAB17} and random\,\cite{AVM11, Cha12} sequence, ballistic spreading to the localization of the wavepacket of the particle has been studied.  One of the mathematically rigorous approaches to understand the asymptotic behavior of the dynamics is to compute the  limit distribution function\,\cite{KLS13, Kon10}. In Ref.\,\cite{MK09}, limit distribution function for the two-period quantum walk using two orthogonal matrices as alternate quantum coin operations has been  computed. Inspite of the important role of quantum interference in the dynamics of the quantum walk it has been shown that the limit distribution of the two-period quantum walk is determined by one of the two quantum coin operations (orthogonal matrix). 

This is an important observation which needs to be explored in more detail to understand the intricacy involved in the dynamics of the periodic quantum walks.  Particularly, when two-period quantum walks is shown to produce the dynamics identical  to the split-step quantum walk\,\cite{ZGS17} which has been used to simulate topological quantum walks, Dirac-cellular automata\,\cite{MC16} and Majorana modes and edge states\,\cite{ZGS17} where both the coin operations play an important role. 

Obtaining the limit density function for the nonorthogonal unitary matrix as the quantum coin operation for the two-period and for other $n-$period quantum walks has been a hard task. Even if one succeeds in meticulously obtaining a limit theorem, it will give us an asymptotic behavior and fails to lay out the way evolutions modulate during each sequence of periodic operations.

In this paper we re-visit the dynamics of the two-period discrete-time quantum walk using nonorthogonal unitary quantum coin operations $\hat{C}(\theta_1)$ and $\hat{C}(\theta_2)$. For the two-period quantum walk with the parameters $\theta_1$ and $\theta_2$ in the coin operations we show that the standard deviation ($\sigma_{\theta_1, \theta_2}$) is the same as the minimum of the standard deviation obtained from the one-period quantum walk with coin operations $\theta_1$ or $\theta_2$, $\sigma_{\theta_1, \theta_2} = \min \{\sigma_{\theta_1}, \sigma_{\theta_2} \}$. Our numerical result is analytically corroborated using the dispersion relation obtained from the continuum limit of the dynamics.  Though the standard deviations are identical,  the spread in position space after $t$ steps is bounded by the $ \pm \abs{t\cos(\theta_1) \cos(\theta_2)}$. And the interference pattern is also clearly distinct. This shows up with the prominent presence of both the parameters $\theta_1$ and $\theta_2$ in the differential form of the dynamics expression. We also show that the bounds we obtained for the two-period quantum walk will hold good for the split-step quantum walk which is defined using two coin operators using $\theta_1$ and $\theta_2$.  Our dispersion relationship approach can be extended to study bounds on the dynamics of three- and higher period quantum walks. Unlike the previous known connection of discrete-time quantum walks with the massless Dirac equation where coin parameter $\theta=0$, here we show the recovery of the gapless (massless) and gapped (massive) Dirac equation with nonzero $\theta$ parameters contributing to the intriguing interference in the dynamics in a totally nonrelativistic situation. We also study the effect of periodic sequence on the entanglement between coin and position space.

In Sec. \ref{Pqw} we will give a basic introduction to the operators that define the evolution of the discrete-time quantum walk. Using that as a basis we will define the periodic quantum walk and present the numerical results for the two-period quantum walk. In Sec.\,\ref{dispers}, we obtain the dispersion relation for the one- and two-period quantum walk and use it to arrive at the bounds on the dynamics of two- and three- and higher- period quantum walks. In Sec.\,\ref{emgDE}, we present the emergence of the Dirac equation from the two-period quantum walk and present the enhancement of entanglement for periodic quantum walks in Sec.\,\ref{EntEn}. We conclude with our remarks in Sec.\,\ref{conc}.

\section{Periodic Quantum Walk}
\label{Pqw}

Dynamics of the one-dimensional discrete-time quantum walk on a particle with two internal degrees of freedom is defined on an  Hilbert
space $\mathcal{H}_w=\mathcal{H}_c \otimes  \mathcal{H}_p$ where the coin Hilbert space $
\mathcal{H}_c  = span \{\ket{\uparrow},  \ket{\downarrow}\}$  and  position Hilbert space $  \mathcal{H}_p  = span
\{\ket{i}\}$, $i  \in \mathbb{Z}$ representing the  number of position states available  to the walker.
The generic initial  state of  the particle, $\ket{\psi}_c$, can be written using a two parameters $\delta$, $\eta$ in the form, 
\begin{equation}
\ket{\psi(\delta, \eta)}_c = \cos(\delta) \ket{0} + e^{-i\eta} \sin(\delta) \ket{1}.
\end{equation}
Each step of the walk evolution is defined by the action of the unitary quantum coin operation followed by the position shift operator.  
The single parameter quantum coin operator which is a non-orthogonal unitary and acts only on the particle space can be written in the form, 
\begin{equation}
\hat{C}(\theta) =
  \begin{bmatrix}
  ~~~  \cos(\theta) & -i \sin(\theta) \\
    -i \sin(\theta) & ~~~ \cos(\theta) 
  \end{bmatrix}.
  \label{qcoin}
\end{equation}
The position shift operator  $\hat{S}$ that  translates the  particle to the left  and/or right conditioned on the internal state of the particle is of the form, 
\begin{equation}
\hat{S}  = \ket{0}\bra{0}
\otimes \sum_{i\in\mathbb{Z}}  \ket{i-1}\bra{i}+\ket{1}\bra{1} \otimes
\sum_{i\in\mathbb{Z}}\ket{i+1}\bra{i}.
\end{equation}
The state of the particle in extended position space after $t$ steps of the homogeneous (one-period) quantum walk is given by applying the operator $\hat{W}=\hat{S}(\hat{C}\otimes I)$ on the initial state of the particle and the position, 
\begin{equation}
\ket{\Psi(t)} = \hat{W}^t \bigg[\ket{\psi}_c \otimes \ket{x=0} \bigg ] = \sum_x \begin{bmatrix} \psi^{\downarrow}_{x, t} \\
\psi^{\uparrow}_{x, t} \end{bmatrix}.
\end{equation}
Probability of finding particle at position and time $(x, t)$ will be
\begin{align}
P(x, t) = \norm{\psi^{\downarrow}_{x, t}}^2 + \norm{\psi^{\uparrow}_{x, t}}^2 .
\end{align}
Using $P(x, t)$ we can compute the standard deviation ($\sigma$) of the probability distribution after $t$ steps of the walk.\\
\begin{figure}
	\begin{center}
	\includegraphics[width=\linewidth]{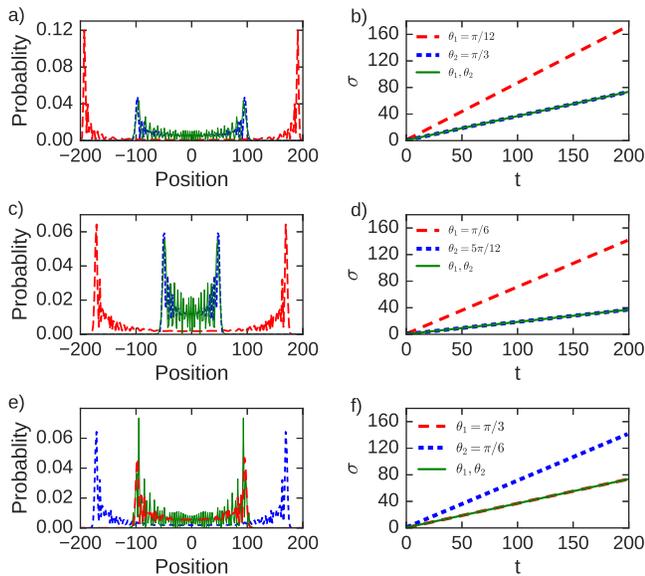}
	\caption{(Color online) Probability distribution after 200 steps of the quantum walk using a different combination of quantum coin operations and a corresponding standard deviation as a function of time. In (a), (c), and (e) we have plotted the probability distribution in position space for both one- and two-period quantum walks. We can notice that the spread of the probability for  the two-period case after $t$ steps is bounded by  $\pm \min  \{  t \abs{\cos(\theta_1)},  t \abs{\cos(\theta_2)} \}$. The standard deviation plot in (b) and (d) shows that $\sigma_{\theta_1, \theta_2} (t)= \sigma_{\theta_2}(t)$  and in (f)   $\sigma_{\theta_1, \theta_2}(t) = \sigma_{\theta_1}(t)$. However, the interference pattern is clearly distinct with prominent oscillations for the two-period case.} 
	\label{fig1}
\end{center}
\end{figure}

\noindent
{\it Two-period quantum walk :} To describe the periodic quantum walk we will use two quantum coin operations $C(\theta_1)$ and $C(\theta_2)$. 
The evolution operator for the $t$ step of the two-period quantum walk will be of the form,
\begin{equation}
[\hat{W}_{\theta_2}\hat{W}_{\theta_1}]^{t/2}.
\end{equation}
For the $n-$period quantum walk the evolution is described using operation $\hat{W}_{\theta_2}$ for every multiple of $n$ steps and $\hat{W}_{\theta_1}$ for all other steps.
We should note that the two-period quantum walk we have defined is a time-dependent periodic evolution but for the localized initial state and evolution operators we have defined it is equivalent to the position-dependent two-period quantum walk. This equivalence should be attributed to the probability distribution which will be zero at the odd (even) position when $t$ is even (odd). But this equivalence will not hold good to any $n-$period quantum walk in general. 

From earlier results we know that the spread of the one-period quantum walk probability distribution using evolution operation $\hat{W}_{\theta}$ is bounded between $-t \cos(\theta)$ and $+t\cos(\theta)$ ($\pm t\cos(\theta)$) and $\sigma \propto t \abs{\cos(\theta)}$\,\cite{NV00, CSL08}. For a two-period walk it looks natural to expect the spread to be bounded somewhere between positions $\pm t \cos(\theta_1)$ and $\pm t \cos(\theta_2)$. But in reality the spread is bounded between $ \pm \min \{t  \abs{\cos(\theta_1)},  t\abs{ \cos(\theta_2)} \}$.

\begin{figure}
	\centering
	\includegraphics[width=\linewidth]{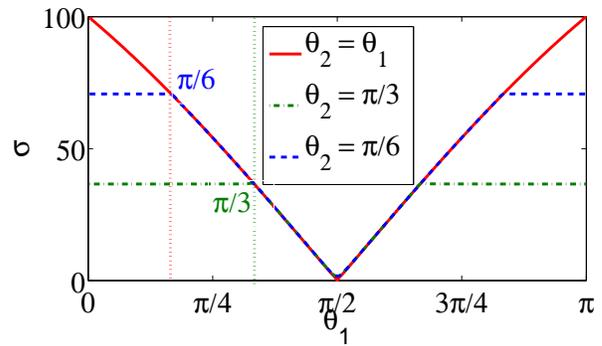}
	\caption{(Color online) Standard deviation ($\sigma$) as a function of $\theta_1$ when $\theta_2$ is fixed. With increase in $\theta_1$ we  notice that $\sigma_{\theta_1, \theta_2}(t) =\min \{ t \abs{\cos(\theta_1)}, t \abs{\cos(\theta_2)} \}$. }
	\label{fig2}
\end{figure}

\begin{figure}
	\centering
	\includegraphics[width=\linewidth]{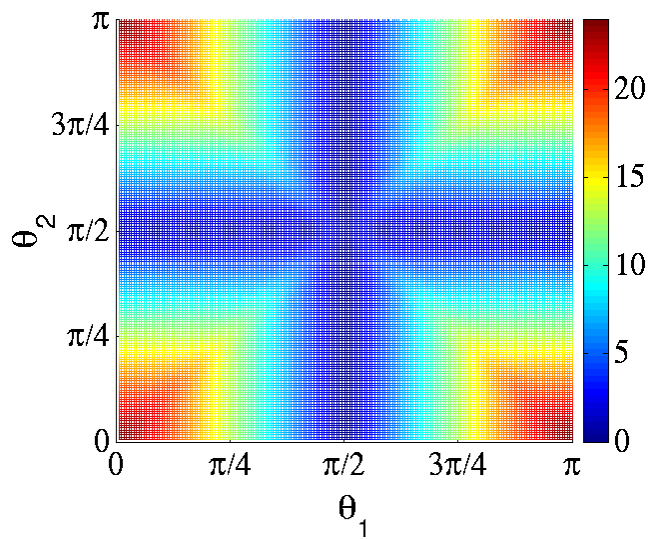}
	\caption{(Color online) Standard deviation as a function of $\theta_1$ and $\theta_2$ after 25 steps of the quantum walk. With increase in both, $\theta_1$ and $\theta_2$ we note that $\sigma_{\theta_1, \theta_2}(t)  =  \min \{ t \abs{\cos(\theta_1)}, t \abs{\cos(\theta_2)} \}$.}
	\label{fig3}
\end{figure}

In Fig.\,\ref{fig1},  the probability distribution and standard deviation ($\sigma$) after 200 steps of the quantum walk using different values of $\theta_1$ and $\theta_2$, separately (one-period) and together in two-period sequence is presented.  We can see that the spread of the probability distribution of the two-period quantum walk $P_{\theta_1, \theta_2}(t)$ is always bounded within the spread of the probability distribution $\min\{P_{\theta_1}(t), P_{\theta_2}(t)\}$ and $\sigma_{\theta_1, \theta_2}(t) = \min\{\sigma_{\theta_1}(t), \sigma_{\theta_2}(t)\}$. But the interference pattern is not identical. In Fig.\,\ref{fig2}, $\sigma_{\theta_1, \theta_2}$ after 100 steps as a function of $\theta_1$ when $\theta_2$ is fixed is presented.  In Fig.\,\ref{fig3}, $\sigma_{\theta_{1}, \theta_2}$  as a function of $\theta_1$ and $\theta_2$ after 25 steps of the quantum walk is shown. Analyzing the dependence of $\sigma$ on the two coin parameters we can note that the $\sigma_{\theta_1, \theta_2}(t)  \propto  \min \{ t \abs{\cos(\theta_1)}, t \abs{\cos(\theta_2)} \}$. 

 In Ref.\,\cite{MK09}, for a combination of orthogonal matrices in the two-period quantum walk, the limit distribution ($L_{1, 2}(X)$) was computed for a specific combination of parameters and shown to be identical to the limit distribution of the quantum walk using the single coin operation, $L_{1, 2}(X) = \min \{L_{1}(X), L_{2}(X) \}$. However, from the probability distribution shown in Fig.\,\ref{fig1}, the interference pattern within the bound is different and the limit distribution function fails to capture that.   To get more insight into the dynamics of the two-period quantum walk and explore the physical significance we will study the dynamic expression at time $t$ and obtain the dispersion relation for it in the continuum limit.



\subsection{Dispersion relation and bounds on spread of wave packet} 
\label{dispers}


\noindent
{\it One-period quantum walk.}
The state of the particle after $t+1$ number of steps of a one-period discrete-time quantum walk can be written as,

\begin{align}
|\Psi(t+1)\rangle  = \sum_{x = -(t+1)}^{t+1} (\psi^\downarrow_{x, t+1} + \psi^\uparrow_{x, t+1})
\end{align}
where the left and right propagating components of the particle is given by, 

\begin{subequations}
\begin{align}
\psi^\downarrow_{x, t+1}  =& \cos(\theta)\psi^ \downarrow_{x+1, t} - i \sin(\theta)\psi^\uparrow_{x+1, t} \\	
\psi^\uparrow_{x, t+1}  =& -i \sin(\theta)\psi^ \downarrow_{x-1, t} + \cos(\theta)\psi^\uparrow_{x-1, t}.
\end{align}
\end{subequations}

This can be written in the matrix form,
\begin{align}
\begin{bmatrix} \nonumber
\psi^\downarrow_{x, t+1} \\
\psi^\uparrow_{x, t+1}
\end{bmatrix}
=&
 \begin{bmatrix}
\cos(\theta) & -i \sin(\theta) \\
0 & 0
\end{bmatrix}
\begin{bmatrix}
\psi^\downarrow_{x+1, t} \\
\psi^\uparrow_{x+1, t}
\end{bmatrix} \\
& + \begin{bmatrix} 
0 & 0 \\
-i \sin(\theta) & \cos(\theta)
\end{bmatrix}
\begin{bmatrix}
\psi^\downarrow_{x-1, t} \\
\psi^\uparrow_{x-1, t}
\end{bmatrix}.
\label{Peq001}
\end{align} 
By adding and subtracting the left-hand side of Eq.\,(\ref{Peq001}) by 
$\begin{bmatrix} \nonumber
\psi^\downarrow_{x, t} \\
\psi^\uparrow_{x, t}
\end{bmatrix}$ and the right-hand side by $\begin{bmatrix} \nonumber
\cos(\theta) & -i \sin(\theta) \\
-i \sin(\theta)  & \cos(\theta)
\end{bmatrix}$ we get a difference operator which can be converted to a differential operator which will result in the differential equation of the form, 
\begin{align}
\label{Peq01}
\frac{\partial}{\partial t} \nonumber
\begin{bmatrix}
\psi^\downarrow_{x,t} \\ \psi^\uparrow_{x,t} 
\end{bmatrix}
= 
& \begin{bmatrix}  
\cos(\theta) & -i \sin(\theta) \\
i \sin(\theta)  & -\cos(\theta)
\end{bmatrix}
\begin{bmatrix}
\frac{\partial \psi^\downarrow_{x, t}}{\partial x} \\
\frac{\partial \psi^\uparrow_{x, t}}{\partial x}
\end{bmatrix} \\
+&\begin{bmatrix}
\cos(\theta)-1 & -i \sin(\theta) \\
-i \sin(\theta) & \cos(\theta)-1
\end{bmatrix}
\begin{bmatrix}
\psi^\downarrow_{x,t} \\ \psi^\uparrow_{x,t} 
\end{bmatrix}.
\end{align}
By reorganizing the preceding expression we get a simultaneous equation of the form,
\begin{subequations}
\label{sim1p}
\begin{align}
\bigg \{ \frac{\partial}{\partial t}  - \cos(\theta) \frac{\partial}{\partial x} -(\cos(\theta) -1) \bigg \} \psi^{\downarrow}_{x, t}  \nonumber \\
+ i \sin(\theta) \bigg \{\frac{\partial}{\partial x} + 1 \bigg \}  \psi^{\uparrow}_{x, t} = 0 \\
\bigg \{ \frac{\partial}{\partial t}  + \cos(\theta) \frac{\partial}{\partial x} -(\cos(\theta) -1) \bigg \} \psi^{\uparrow}_{x, t}  \nonumber \\
+ i \sin(\theta) \bigg \{\frac{\partial}{\partial x} - 1 \bigg \}  \psi^{\downarrow}_{x, t} = 0.
\end{align}
\end{subequations}
For the above expression governing the dynamics of each step of the one-period quantum walk in the continuum limit, we can seek a Fourier-mode wave like solution of the form 
\begin{equation}
\psi_{x, t} = e^{i(kx -\omega t)},
\end{equation}
where $\omega$ is the wave frequency and $k$ is the wave number. Upon substitution into the real part of Eq.\,(\ref{sim1p}) we get,
\begin{align}
\omega = \mp  k \cos(\theta) + i[\cos(\theta)-1] 
\label{Peqdis}
\end{align}
and the group velocity will be
\begin{align}
v^g_1 = \frac{d \omega}{d k} = \mp \cos(\theta).
\end{align}
From this we can say that the wavepacket spreads at a rate of $\cos(\theta)$ during each step of the quantum walk and after $t$ steps the spread will be between $\pm t \cos(\theta)$. Though we have used only one form of the quantum coin operation with complex elements in it, the group velocity will be $\propto \cos(\theta)$ even when a most generic unitary operator is used as a quantum coin operation\,\cite{Cha12}. 

\noindent{\it Two-period quantum walk :}
For the two-period quantum walk the evolution is driven by two quantum coin operations $\hat{C}(\theta_1)$ and $\hat{C}(\theta_2)$. First, we will write the state at position $x$ and time $t+1$, $\psi^{\downarrow(\uparrow)}_{x, t+1}$ as a component of $\theta_2$ at time $t$,
\begin{subequations}
\begin{align}
\psi^\downarrow_{x, t+1} & = \cos(\theta_2)\psi^ 	\downarrow_{x+1, t} - i \sin(\theta_2)\psi^\uparrow_{x+1, t}   \\	
\psi^\uparrow_{x, t+1} & = -i \sin(\theta_2)\psi^ 	\downarrow_{x-1, t} + \cos(\theta_2)\psi^\uparrow_{x-1, t}.
\end{align} 
\label{ct2}
\end{subequations}
In the preceding expression, dependency of the state $\psi^{\downarrow (\uparrow)}_{x, t+1}$ on the coin parameter $\theta_1$ can be obtained by writing the state $\psi^{\downarrow(\uparrow)}_{x \pm 1, t}$
as component of $\theta_1$ at time $(t-1)$, 
\begin{subequations}
\begin{align}
\psi^\downarrow_{x+1, t} &=   \cos(\theta_1)\psi^\downarrow_{x+2, t-1} - i \sin(\theta_1)\psi^\uparrow_{x+2, t-1}  \\
\psi^\uparrow_{x+1, t} &= -i \sin(\theta_1)\psi^\downarrow_{x, t-1} + \cos(\theta_1) \psi^\uparrow_{x, t-1}   \\
\psi^\downarrow_{x-1, t} &=  \cos(\theta_1)\psi^\downarrow_{x, t-1} - i \sin(\theta_1)\psi^\uparrow_{x, t-1}  \\
\psi^\uparrow_{x-1, t} &= -i \sin(\theta_1)\psi^\downarrow_{x-2, t-1} + \cos(\theta_1)\psi^\uparrow_{x-2, t-1}
\end{align}
\label{ct1}
\end{subequations}
Now, substituting Eq.\,(\ref{ct1}) into Eq.\,(\ref{ct2}) we obtain, 
\begin{subequations}
\begin{align}
\psi^\downarrow_{x, t+1} = \cos(\theta_2)[\cos(\theta_1)\psi^\downarrow_{x+2, t-1} - i \sin(\theta_1)\psi^\uparrow_{x+2, t-1}] \nonumber\\ 
- i \sin(\theta_2) [-i \sin(\theta_1)\psi^\downarrow_{x, t-1}+ \cos(\theta_1)\psi^\uparrow_{x, t-1}]   \\
\psi^\uparrow_{x, t+1} = -i  \sin(\theta_2)[\cos(\theta_1)\psi^\downarrow_{x, t-1} - i  \sin(\theta_1)\psi^\uparrow_{x, t-1}] \nonumber \\ 
+  \cos(\theta_2) [-i \sin(\theta_1)\psi^\downarrow_{x-2, t-1}+ \cos(\theta_1)\psi^\uparrow_{x-2, t-1}].
\end{align}
\end{subequations}
\begin{figure}
	\includegraphics[width=\linewidth]{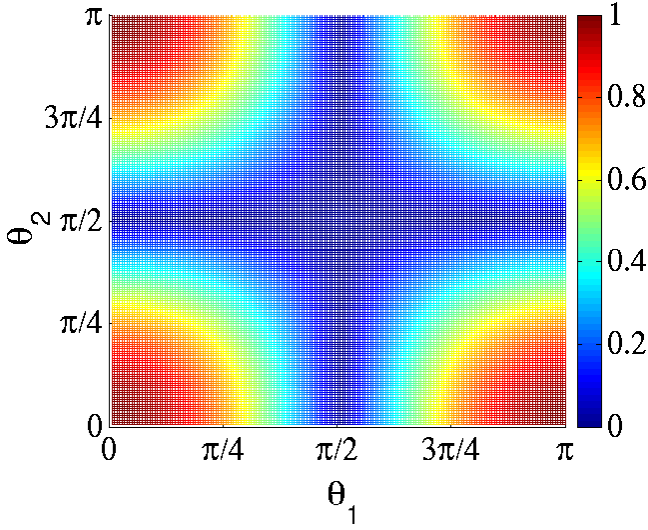}
	\caption{(Color online) Group velocity obtained from the dispersion relation as a function of $\theta_1$ and $\theta_2$ for the two-period quantum walk.  The group velocity obtained in the continuum limit of evolution for each step of the walk when multiplied by the number of steps of the walk it matches with the overall pattern of standard deviation obtained in discrete evolution of the walk. } 
	\label{GroupVel}
\end{figure}
Without loosing any generic feature in the preceding evolution expression we can replace $t$ with $t+1$. After that we can effectively reduce the two-step evolution expression using coins with parameters $\theta_1$ and $\theta_2$ to a combined single-step evolution expression by replacing $x\pm2$ in the right-hand side by $x\pm1$ and $t+2$ in the left-hand side by $t+1$. This will result in 
\begin{subequations}
\begin{align}
\psi^\downarrow_{x, t+1} = \nonumber \cos(\theta_2)[\cos(\theta_1)\psi^\downarrow_{x+1, t} - i \sin(\theta_1)\psi^\uparrow_{x+1, t}]\\ - i \sin(\theta_2) 
[-i \sin(\theta_1)\psi^\downarrow_{x, t}+ \cos(\theta_1)\psi^\uparrow_{x, t}] \\
\psi^\uparrow_{x, t+1} = \nonumber -i \sin(\theta_2)[\cos(\theta_1)\psi^\downarrow_{x, t} - i \sin(\theta_1)\psi^\uparrow_{x, t}]\\ +  \cos(\theta_2) 
[-i \sin(\theta_1)\psi^\downarrow_{x-1, t}+ \cos(\theta_1)\psi^\uparrow_{x-1, t}]. 
\end{align}
\label{evol1}
\end{subequations}
In the matrix form this can be written as,
\begin{align}
\begin{bmatrix} \nonumber
\psi^\downarrow_{x, t+1} \\
\psi^\uparrow_{x, t+1}
\end{bmatrix}
=
& \begin{bmatrix} \nonumber
-\sin(\theta_2)\sin(\theta_1) & 
-i \sin(\theta_2)\cos(\theta_1) \\
-i \sin(\theta_2)\cos(\theta_1) & 
-\sin(\theta_2)\sin(\theta_1)   
\end{bmatrix}
\begin{bmatrix}
\psi^\downarrow_{x, t} \\
\psi^\uparrow_{x, t}
\end{bmatrix} \\
+&
\begin{bmatrix} \nonumber
0 & 0 \\
-i \cos(\theta_2)\sin(\theta_1) & \cos(\theta_2)\cos(\theta_1)
\end{bmatrix}
\begin{bmatrix}
\psi^\downarrow_{x-1, t} \\
\psi^\uparrow_{x-1, t}
\end{bmatrix} \\
+&
\begin{bmatrix}
\cos(\theta_2)\cos(\theta_1) & -i \sin(\theta_1)\cos(\theta_2) \\
0 & 0
\end{bmatrix}
\begin{bmatrix}
\psi^\downarrow_{x+1, t} \\
\psi^\uparrow_{x+1, t}
\end{bmatrix}.
\label{de1}
\end{align} 
By adding and subtracting the left-hand side of Eq.\,(\ref{de1}) by $\begin{bmatrix} \nonumber
\psi^\downarrow_{x, t} \\
\psi^\uparrow_{x, t}
\end{bmatrix}$ and the right-hand side by $\begin{bmatrix} \nonumber
\cos(\theta_2)\cos(\theta_1) & -i \sin(\theta_1)\cos(\theta_2) \\
-i \sin(\theta_1) \cos(\theta_2) & \cos(\theta_2)\cos(\theta_1)
\end{bmatrix}$ we get a difference operator which can be converted to a differential operator which will result in the differential equation of the form, 

 \begin{figure}
	\centering
	\includegraphics[width=\linewidth]{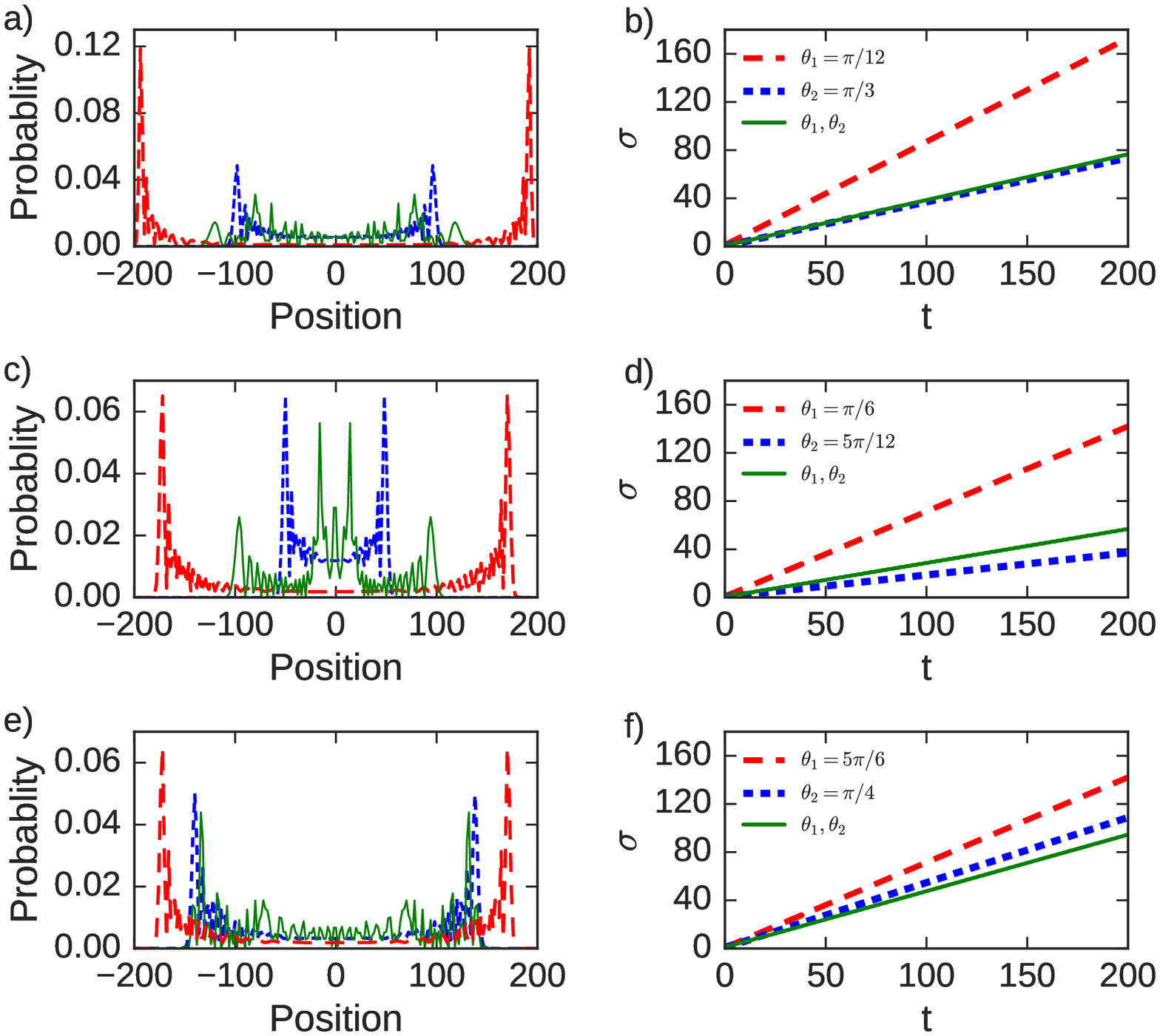}
	\caption{(Color online) Probability distribution after 200 steps of the quantum walk for a different combination of quantum coin operations and a corresponding standard deviation as a function of time. In (a), (c), and (e) we have plotted the probability distribution in position space for both the one- and three-period quantum walks. We can notice that the spread of the probability for the three-period case after $t$ steps is always lower than $\pm \max  \{  t \abs{\cos(\theta_1)},  t \abs{\cos(\theta_2)} \}$ but not bounded by the minimum of the two as it was for the two-period quantum walk. The standard deviation plot in (b), (d), and (f) shows that $\sigma_{\theta_1, \theta_2} (t)$  will be around $\min \{\sigma_{\theta_1}(t), \sigma_{\theta_2}(t) \}$. }
	\label{fig5}
\end{figure}
\begin{align}
\frac{\partial}{\partial t} \nonumber
\begin{bmatrix}
\psi^\downarrow_{x,t} \\ \psi^\uparrow_{x,t} 
\end{bmatrix}
= 
& \cos(\theta_2) \begin{bmatrix}  
\cos(\theta_1) & -i \sin(\theta_1) \\
i \sin(\theta_1)  & -\cos(\theta_1)
\end{bmatrix}
\begin{bmatrix}
\frac{\partial \psi^\downarrow_{x, t}}{\partial x} \\
\frac{\partial \psi^\uparrow_{x, t}}{\partial x}
\end{bmatrix} \\
+&\begin{bmatrix}
\cos(\theta_1+\theta_2)-1 & -i \sin(\theta_1+\theta_2) \\
-i \sin(\theta_1+\theta_2) & \cos(\theta_1+\theta_2)-1
\end{bmatrix}
\begin{bmatrix}
\psi^\downarrow_{x,t} \\ \psi^\uparrow_{x,t} 
\end{bmatrix}.
\label{Peq}
\end{align}
The preceding matrix representation can be reorganized and written as a simultaneous equations, 
\begin{subequations}
\begin{align}
\bigg \{\frac{\partial}{\partial t} - \cos(\theta_2)\cos(\theta_1)\frac{\partial}{\partial x} - [\cos(\theta_1+\theta_2)-1] \bigg \} \psi^\downarrow_{x,t} \nonumber \\
+i \bigg \{\sin(\theta_1)\cos(\theta_2)\frac{\partial}{\partial x} +\sin(\theta_1 +\theta_2) \bigg \} \psi^\uparrow_{x,t} =0 \\
\bigg \{\frac{\partial}{\partial t} + \cos(\theta_2)\cos(\theta_1)\frac{\partial}{\partial x} - [\cos(\theta_1+\theta_2)-1] \bigg \} \psi^\uparrow_{x,t} \nonumber \\
-i \bigg \{\sin(\theta_1)\cos(\theta_2)\frac{\partial}{\partial x} -\sin(\theta_1 +\theta_2) \bigg \} \psi^\downarrow_{x,t} =0.
\end{align}
\label{Peq1}
\end{subequations}
For the above expression effectively governing the dynamics of the two-period quantum walk in the continuum limit, we can seek a Fourier-mode wave like solution of the form 
$\psi_{x, t} = e^{i(kx -\omega t)}$.
Upon substitution into the real part of Eq.\,(\ref{Peq1}) we get,
\begin{align}
\omega = \mp  k \cos(\theta_2)\cos(\theta_1) + i[\cos(\theta_1+\theta_2)-1] 
\label{Peqdis}
\end{align}
and the group velocity will be
\begin{align}
v^g_2 = \frac{d \omega}{d k} = \mp \cos(\theta_2)\cos(\theta_1).
\label{grvel}
\end{align}
In Fig.\,\ref{GroupVel} we have plotted group velocity for the two-period quantum walk, $v^g_2(\theta_1, \theta_2)$. This gives an effective displacement of the wave packet  for each step of the two-period quantum walk when the two-step evolution using $\theta_1$ and $\theta_2$ is combined to one effective step evolution. Comparing Figs.\,\ref{GroupVel} and \,\ref{fig3}, $v_g$ and  $\sigma$ as a function of $\theta_1$ and $\theta_2$ we can see an identical pattern and only when the dominance of  one $\theta$ over the other happens, the transition is smooth for $v_g$. This is due to the continuum approximation we made in the analytics. 

From the expression for group velocity, Eq.\,(\ref{grvel}), we can infer that 
\begin{align}
\abs{v^g_2} \leq  \min \{ \abs {\cos(\theta_1)}, \abs {\cos(\theta_2)} \}.
\end{align}
Therefore, the bound on the group velocity sets the bound on the standard deviation, $\sigma (t) \propto t \abs{v^g_2}$.  This bound on the group velocity and standard deviation corroborates with the bounds we  obtained from the numerical analysis.

\begin{figure}
	\includegraphics[width=\linewidth]{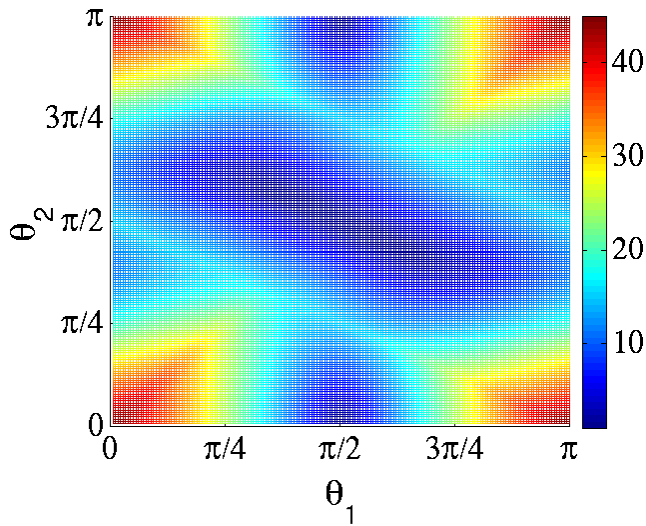}
	\caption{(Color online) Standard deviation as function of $\theta_1$ and $\theta_2$ after 45 steps of the three-period quantum walk.  Except for $\theta$'s where  $\abs{\cos(\theta_1)} \approx  \abs{\cos(\theta_2)}$ and close to unity,  the standard deviation is very low. This can be attributed to multiple peaks in the distribution where peaks with higher probability are closer to the origin. } 
	\label{fig6}
\end{figure}
\noindent
{\it Three- and $n-$period quantum walk :} 
First three step of the three-period quantum walk using two quantum coin operations $\hat{C}(\theta_1)$ and $\hat{C}(\theta_2)$ is implemented with the evolution operator in sequence, 
\begin{equation}
\hat{W}_{3P} = \hat{W}_{\theta_2}\hat{W}_{\theta_1}\hat{W}_{\theta_1}.
\end{equation}
In Fig.\,\ref{fig5}, the probability distribution after 200 steps of  the three-period quantum walk is presented and the spread of the probability after $t$ steps is always lower than $\pm \max  \{  t \abs{\cos(\theta_1)},  t \abs{\cos(\theta_2)} \}$ but not bounded by the minimum of the two as it was for two-period quantum walk. The standard deviation plot in Figs.\,\ref{fig5}(b), \ref{fig5}(d), and \ref{fig5}(f) shows that $\sigma_{\theta_1, \theta_2} (t)$  will be around $\min \{\sigma_{\theta_1}(t), \sigma_{\theta_2}(t) \}$ and does not match explicitly.   In Fig.\,\ref{fig6}, the standard deviation as a function of $\theta_1$ and $\theta_2$ after 45 steps of the three-period quantum walk is shown.  Except for the evolution parameter where  $\abs{\cos(\theta_1)} \approx  \abs{\cos(\theta_2)}$ and close to unity,  the standard deviation is very low. This can be attributed to multiple peaks in the distribution where peaks with higher probability are closer to origin. 

Unlike the two-period case where only two peaks were seen in the probability distribution, multiple peaks can emerge in the three- and $n-$period quantum walks (see Fig.\,\ref{fig8}). This can result in a mismatch between the linear scaling of group velocity with the standard deviation.  However, group velocity can give us a definite bound on the maximum spread of the probability distribution in position space for three- and $n-$period quantum walks.
\begin{figure}
	\includegraphics[width=\linewidth]{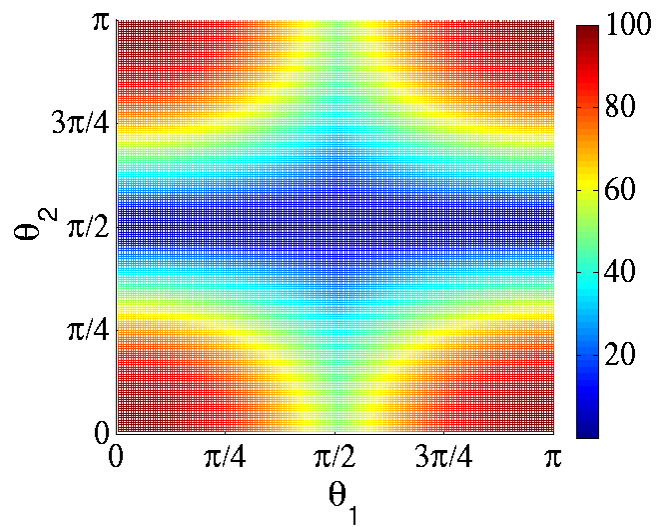}
	\caption{(Color online) Spread of the probability distribution in position space after 100 steps of the three-period quantum walk as a function of $\theta_1$ and $\theta_2$.  This bound on the spread is obtained from the maximum of group velocity $v^g_3$.} 
	\label{fig7}
\end{figure}
\begin{figure}
	\centering
	\includegraphics[width=\linewidth]{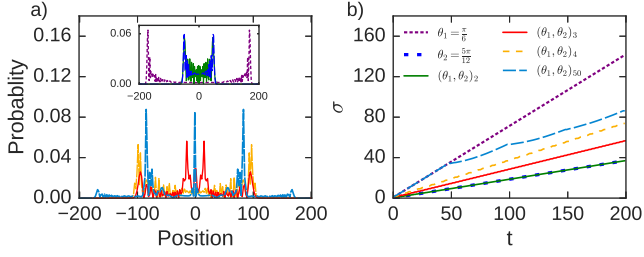}
	\caption{(Color online) Probability distribution and standard deviation after 200 steps of the $n-$period quantum walk. In (a) the probability distribution for three-period,  four-period, and fifty-period quantum walks is shown. The inset shows the position probability distribution for the two-period and when the coin is homogeneous (one-period) with the coin parameters $\theta_1$ and $\theta_2$ . The standard deviation (b) shows only the two-period evolutions is bounded by $\theta_1$; for the three-period and four-period evolutions it is bounded between $\theta_2$ and $\theta_1$.  We can verify that the spread in probability distribution is bounded by a maximum of group velocity for all $n-$period quantum walks.}
	\label{fig8}
\end{figure}

The evolution operator for the first three steps of the three-period walk can be re-written as,
\begin{equation}
\hat{W}_{3P} = \hat{W}_{2P}\hat{W}_{\theta_1} ,
\end{equation}
where $\hat{W}_{2P}$ represent the two-period operator sequence for which we already know the dispersion relation and $v^g_2$ [Eq.\,(\ref{grvel})] when it is treated as an effective one step evolution.  Extrapolating $v^g_1$  and $v^g_2$ from one-period and two-period quantum walks we can write the group velocity for the three-period walk in the form,
\begin{equation}
v^g_3 = \frac{\pm( v^g_1 + v^g_2)}{2} = \pm \frac{1}{2}\bigg [ \cos(\theta_1) \pm \cos(\theta_1) \cos(\theta_2) \bigg ].
\end{equation}
For any given values of $\theta$'s, we can get multiple valid value for $v^g_3$ . This can be interpreted as the wave packet simultaneously evolving with different $v^g_3$ resulting in multiple peaks in the probability distribution. Among the possible values for $v^g_3$  the contribution for a maximum spread in position space will be from, 
\begin{equation}
\max \abs{ v^g_3} = \frac{1}{2} \bigg [ \abs{\cos(\theta_1)} + \abs{\cos(\theta_1) \cos(\theta_2)} \bigg ].
\end{equation}
From the preceding expression we can conclude that the bound on the spread of the wave packet in position space after the $t$ step of the three-period walk will be,
\begin{equation}
\pm t ~\max \abs{ v^g_3}  = \frac{\pm t}{2} \bigg [ \abs{\cos(\theta_1)} + \abs{\cos(\theta_1) \cos(\theta_2)} \bigg ].
\label{maxgv}
\end{equation}
In Fig.\,\ref{fig7}, bounds on the spread of the probability distribution in position space after 100 steps of the three-period quantum walk as a function of $\theta_1$ and $\theta_2$ is shown.  This bound on the spread is obtained from the maximum of group velocity $v^g_3$. By substituting finite values for $\theta_1$ and $\theta_2$ into Eq.\,(\ref{maxgv}) we can confirm that the bounds we get from a maximum of group velocity matches with the maximum range of spread of probability distribution obtained from numerical evolution [Fig.\,\ref{fig5}(a), Fig.\,\ref{fig5}(c), Fig.\,\ref{fig5}(e) and Fig.\,\ref{fig8}].

For the $n-$period quantum walk, the spread of the probability distribution will be bounded by,
\begin{equation}
\pm t ~\max \abs{ v^g_n}  = \frac{\pm t}{(n-1)} \bigg [(n-2)\abs{\cos(\theta_1)} + \abs{\cos(\theta_1) \cos(\theta_2)} \bigg ].
\label{maxgv1}
\end{equation}
In Fig.\,\ref{fig8}, the probability distribution and standard deviation after 200 steps of $n-$period quantum walks is shown. We can verify that the spread in probability distribution is bounded by a maximum of group velocity for all $n-$period quantum walks.
\begin{figure}
	\centering
	\includegraphics[width=\linewidth]{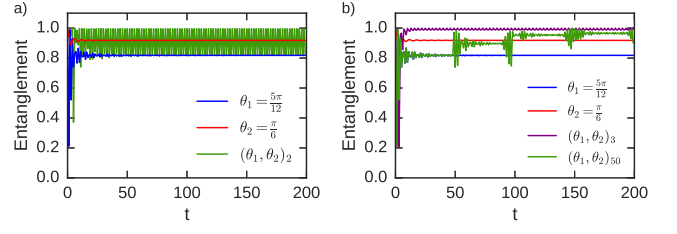}
	\caption{(Color online) Entanglement between the particle and position for 200 steps for one-, two-, three- and fifty-period quantum walks. For the two-period quantum walk (a), in contrast to the standard deviation, the mean value of entanglement is bounded around the maximum of the two on-period quantum walk. For the three-period quantum walk, entanglement reaches a maximum possible value and from the larger $n-$ period quantum walk we can see how the enhancement happens when the quantum coin operation with $\theta_2$  is introduced periodically.}
	\label{ent1}
\end{figure}


  \section{Two-period quantum walk, split-step quantum and Dirac equation}
  \label{emgDE}

The split-step quantum walk was first introduced to define the topological quantum walk\,\cite{KRB10} and was shown to simulate Dirac cellular automata\,\cite{MC16}. Recently, the decomposed form of the split-step quantum walk was shown to be equivalent to the two-period quantum walk and simulate Majorana modes and edge states\,\cite{ZGS17}. In this section, staring from the split-step quantum walk we arrive at the differential equation form of the evolution equation which is equivalent to the two-period quantum walk evolution equation. From this we can establish that all bounds applicable to the two-period quantum walk will hold good for the split-step quantum walk and equivalent form of Dirac equations. 

Each step of the split-step quantum walk is a composition of two half step evolutions,
\begin{equation}
\hat{W_{ss}} = \hat{S}_{+}(\hat{C}(\theta_2) \otimes I) \hat{S}_{-} (\hat{C}(\theta_1)\otimes I),
\end{equation}
where $\hat{C}(\theta_1)$ and $\hat{C}(\theta_2)$ are the quantum coin operation and we will define it in the same form as Eq.\,(\ref{qcoin}). The position shift operators are defined as,
\begin{subequations}
\begin{align}
\hat{S}_{-} =  \ket{0}\bra{0} \otimes \sum_{i\in\mathbb{Z}}  \ket{i-1}\bra{i}+\ket{1}\bra{1} \otimes
\sum_{i\in\mathbb{Z}}\ket{i}\bra{i}\\
\hat{S}_{+} =  \ket{0}\bra{0} \otimes \sum_{i\in\mathbb{Z}}  \ket{i}\bra{i}+\ket{1}\bra{1} \otimes
\sum_{i\in\mathbb{Z}}\ket{i+1}\bra{i}.
\end{align}
\end{subequations}
The state at any position $x$  and time $t+1$ after the operation of $\hat{W}_{ss}$ at time $t$ will be $\psi_{x, t+1} = \psi^{\downarrow}_{x, t+1}  + \psi^{\uparrow}_{x, t+1}$, where
\begin{subequations}
\begin{align}
\psi^\downarrow_{x, t+1} = \cos(\theta_2)[\cos(\theta_1)\psi^\downarrow_{x+1, t} - i \sin(\theta_1)\psi^\uparrow_{x+1, t}] \nonumber\\ 
- i \sin(\theta_2) [-i \sin(\theta_1)\psi^\downarrow_{x, t}+ \cos(\theta_1)\psi^\uparrow_{x, t}]   \\
\psi^\uparrow_{x, t+1} = -i  \sin(\theta_2)[\cos(\theta_1)\psi^\downarrow_{x, t} - i  \sin(\theta_1)\psi^\uparrow_{x, t}] \nonumber \\ 
+  \cos(\theta_2) [-i \sin(\theta_1)\psi^\downarrow_{x-1, t}+ \cos(\theta_1)\psi^\uparrow_{x-1, t}].
\end{align}
\label{deqA}
\end{subequations}
The preceding expression is identical to Eq.\,(\ref{evol1}) which we have obtained for the two-period quantum walk. Therefore, the differential equation form of the evolution will be the same as Eq.\,(\ref{Peq}). By controlling the parameters $\theta_1$ and $\theta_2$ we can arrive at the Dirac equations  for massless and massive particles. 
\begin{enumerate}
\item Multiplying Eq.\,(\ref{Peq}) by $i\hbar$ and setting $\theta_1 = 0$ and $\theta_2$ to a small value (mass of sub-atomic particles) we recover Dirac equation in the form,
\begin{align}
 i \hbar\Bigg [ \frac{\partial}{\partial t} 
 - \bigg (1-\frac{\theta^2_2}{2} \bigg )\begin{bmatrix}  
1 & 0 \\
0  & -1
\end{bmatrix} \frac{\partial}{\partial x}
+ i\theta_2 \begin{bmatrix}
0 & 1  \\
1  & 0
\end{bmatrix} \Bigg ]
\begin{bmatrix}
\psi^\downarrow_{x,t} \\ \psi^\uparrow_{x,t} 
\end{bmatrix} \approx 0.
\label{De1}
\end{align}
\item  By choosing $\theta_1$ and $\theta_2$ such that $\cos(\theta_1 + \theta_2)=1$ in Eq.\,(\ref{Peq}), and multiplying by $i\hbar$ we get an expression identical to Dirac equation of massless particle, 
\begin{align}
 i\hbar \Bigg [ \frac{\partial}{\partial t}  -
 \cos(\theta_2)
&\begin{bmatrix}
\cos(\theta_1) &  -i\sin(\theta_1) \\
i\sin(\theta_1)  & -\cos(\theta_1)
\end{bmatrix}
 \frac{\partial }{\partial x}  \Bigg ]
\begin{bmatrix} 
\psi^\downarrow_{x,t} \\ \psi^\uparrow_{x,t} 
\end{bmatrix} =  0.
\label{De2}
\end{align}
Here the co-efficient of the position derivative is a more general Hermitian matrix which depicts the oscillation of the spin (eigen state)  during the dynamics.

\item By choosing $\theta_1$ to be extremely small and corresponding $\theta_2$ such that $\cos(\theta_1 + \theta_2)=1$ in Eq.\,(\ref{Peq}), and multiplying by $i\hbar$ we get the Dirac equation in the form, 
\begin{align}
i\hbar \Bigg [ \frac{\partial}{\partial t} 
-   \ \cos(\theta_2)
&\begin{bmatrix}
1 & 0 \\
0  & -1
\end{bmatrix}
 \frac{\partial }{\partial x}  \Bigg ]
\begin{bmatrix} 
\psi^\downarrow_{x,t} \\ \psi^\uparrow_{x,t} 
\end{bmatrix} \approx 0.
\label{De3}
\end{align}
\end{enumerate}
In Ref.\,\cite{MC16}, it was shown that $\theta_1 =0$ and small value of $\theta_2$ is required to recover Dirac cellular automata from split-step quantum walk and both $\theta_1 = \theta_2 = 0$ to recover massless Dirac equation. Here, we have shown the other possible configurations of non-zero $\theta$ values where we can recover the massless Dirac equation.  From bounds on the two-period quantum walk (equivalently the split-step walk) we can imply that the spread of the wave packet for the  massive and massless, that is, the gapped and gapless Dirac equation of the form, Eq.\,(\ref{De1}) and Eq.\,(\ref{De3}), respectively  is bounded by the parameter $\theta_2$. The spread will be very wide for the former and small for the latter (remaining around the origin).  For the massless Dirac equation with general the Hermitian matrix, Eq.\,(\ref{De2}),  the spread will be bounded by $\min \{ \cos(\theta_1), \cos(\theta_2) \}$.

\section{Entanglement in periodic quantum walks}
\label{EntEn}

Entanglement of the particle with position during quantum walk evolution has been reported in many earlier studies. Entanglement during the temporal disordered (spatial disorder) quantum walk is reported to be higher (lower) than the homogenous (one-period) quantum walk\,\cite{Cha12}. In the homogenous quantum walk, mean value of entanglement generated is independent of the initial state of the particle. But in the split-step quantum walk, the dependence of mean value of entanglement is prominently visible\,\cite{MC16}.   
Therefore, for the two-period quantum walk, entanglement behavior will be identical to the one reported in Ref.\,\cite{MC16}. In this section we will see how the entanglement manifests and reaches maximum value for the $n-$period quantum walk. 

As we have considered only a pure quantum state evolution in this study, we will use the partial entropy as a measure of entanglement, which is enough to give correct measure of entanglement for the pure state evolution with unitary operators. We will first take the partial trace with respect to  $\mathcal{H}_p$-space (position space) of the time evolved state 
       = $Tr_p(\rho(t)) := \rho_c(t).$
       Then according to our measure the entanglement at time $t$ is given by, 
       \begin{align}
       -Tr_c[\rho_c(t) \log_2 \{\rho_c(t)\}],
       \end{align}
where the suffix {\it c} represents the coin space. 

In Fig.~\ref{ent1}, we present the entanglement between the particle and position space as a function of time for one-,  two-, and $n-$period quantum walks. For the two-period quantum walk, in contrast to standard deviation, the mean value of entanglement is bounded around the maximum of the two one-period quantum walks. For the three-period quantum walk, entanglement reaches a maximum value, higher than the entanglement due to both, one-period quantum walks is seen. This is also in contrast to the way the spread in position space and standard deviation decreases for periodic quantum walks. For the higher period quantum walk we can see that the change of coin induces the increase in entanglement. \\
\\

\vskip 0.5in
\section{Conclusion}  
\label{conc}

In this work we have presented the dynamics of the time-dependent periodic quantum walk. In particular, we have shown the way the probability distribution spreads, standard deviation increases and entanglement varies for the periodic quantum walk; and we have shown the way they are bounded when compared with the dynamics properties of the homogeneous (single coin driven) quantum walk. For the two-period quantum walk with the parameters $\theta_1$ and $\theta_2$ in the coin operations we show that $\sigma_{\theta_1, \theta_2} = \min \{\sigma_{\theta_1}, \sigma_{\theta_2} \} \propto \min\{ t \abs{\cos(\theta_1)}, t \abs{\cos(\theta_2)}$. 
Our numerical results were corroborated with analytical analysis from the dispersion relation of the two-period quantum walk. Re-visiting the split-step quantum walk dynamics we have also shown that all the bounds we have presented for the two-period quantum walk will be identical to the split-step quantum walk.
Unlike the computing limit density function which is meticulously hard, we have used the dispersion relation from one-period and two-period quantum walk to understand the bounds on the spread of the wave packet for the $n-$period quantum walk, $\frac{\pm t}{(n-1)} \bigg [(n-2)\abs{\cos(\theta_1)} + \abs{\cos(\theta_1) \cos(\theta_2)} \bigg ]$. By re-visiting the connection of quantum walks with the Dirac equation, we have shown the configuration of periodic quantum walk evolution which can recover the Dirac equation for both, massive and massless particles with the nonzero coin parameter $\theta$.  Thus, the evolution configuration that results is the emergence of the gapless and gapped Dirac equations. This can contribute to quantum simulation of dynamics in Dirac materials. We also showed that the periodic sequence will enhance the entanglement between the coin and position space in the quantum walk dynamics. 

Depending on the convenience of the experimental system, either the, two-period or split-step quantum walk can be used for quantum simulations of various low-energy and higher energy particle dynamics defined by Dirac equations. The bounds we have presented will further help to understand the transition from the diffusive to the localized state.


\vskip 0.2in
\noindent
{\bf Acknowledgment:}\\
\\
\noindent
CMC and NPK would like to thank Department of Science and Technology, Government of India for the Ramanujan Fellowship grant No.:SB/S2/RJN-192/2014. CMC and RL would also acknowledge the support from US Army Research Laboratory.

\end{document}